\documentclass[11pt,british]{article}
\usepackage[T1]{fontenc}
\usepackage[utf8]{inputenc}
\usepackage[authoryear]{natbib}

\makeatletter

\date{}

\makeatother

\usepackage{babel}
\begin{document}
\title{AI Mimicry and Human Dignity \\
	\large Chatbot Use as a Violation of Self-Respect}
\author{Jan-Willem van der Rijt, Dimitri Coelho Mollo \& Bram Vaassen\\
Umeå University\\
\emph{Preprint - February 2025}}
\maketitle
\begin{abstract}
This paper investigates how human interactions with AI-powered chatbots
may offend human dignity. Current chatbots, driven by large language
models (LLMs), mimic human linguistic behaviour but lack the moral
and rational capacities essential for genuine interpersonal respect.
Human beings are prone to anthropomorphise chatbots---indeed, chatbots
appear to be deliberately designed to elicit that response. As a result,
human beings’ behaviour toward chatbots often resembles behaviours
typical of interaction between moral agents. Drawing on a second-personal,
relational account of dignity, we argue that interacting with chatbots
in this way is incompatible with the dignity of users. We show that,
since second-personal respect is premised on reciprocal recognition
of second-personal moral authority, behaving towards chatbots in ways
that convey second-personal respect is bound to misfire in morally
problematic ways, given the lack of reciprocity. Consequently, such
chatbot interactions amount to subtle but significant violations of
self-respect---the respect we are dutybound to show for our own dignity.
We illustrate this by discussing four actual chatbot use cases (information
retrieval, customer service, advising, and companionship), and propound
that the increasing societal pressure to engage in such interactions
with chatbots poses a hitherto underappreciated threat to human dignity.
\end{abstract}

\section*{Introduction}

Interactions with chatbots are increasingly common. Some such interactions
are completely voluntary, but others are less so, as when a client
contacts a helpdesk that covertly uses a chatbot rather than a human
employee to provide customer assistance. Current chatbots raise a
variety of ethical issues. Previous research identified potential
problems with the exploitation of workers and data, copyright, representativity
and power concentration, sustainability and environmental cost, as
well as discrimination and bias \citep{Weidinger2021,Gabriel2024}.
In this paper, we argue that current chatbots pose an additional,
currently underappreciated ethical threat: interacting with them can,
and often does, offend human dignity. 

It is widely held that special demands and obligations, especially
related to respect, apply to beings endowed with dignity, who have
a distinctive moral standing.\footnote{There is, however, much less agreement on what grounds such special
standing, a debate we cannot go into here (but see e.g. \citealt{Debes2017,Duewell2014,McCrudden2014,Rosen2012,Schaber2012,Pfordten2023}
for overviews).} We build on a second-personal, relational approach to dignity \citep{Darwall2006,Zylberman2017},
according to which the demands and obligations tied to dignity involve
not only appropriately respecting other beings endowed with it, but
also oneself. We argue that this self-respect includes a demand that
we not treat beings that lack the relevant moral standing in the same
way as we treat our equals. In other words, displaying the kind of
respect that attaches to beings with dignity toward beings that lack
it, intentionally or else, constitutes a form of self-debasement,
offending one’s dignity.
\newline
We submit that this is what happens in various interactions with current
chatbots. Users are invited to, lured into, or socially and practically
pressured---sometimes even forced---to treat the chatbot as if it
were an equal; as something with beliefs, desires, opinions, and moral
capacities, capable of providing its own pieces of advice and practical
guidance, as well as reasons for its outputs. Current chatbots, however,
lack these capacities. When we treat chatbots as our equals in this
way, we thus offend our own dignity.

The paper is structured as follows. In section 1, we present some
chatbot use cases pertinent to our analysis. We provide a brief primer
on current chatbots in Section 2. Section 3 introduces the second-personal,
relational approach to dignity we endorse in this paper, and how it
accounts for offences to dignity. We go back to the chatbot use cases
in section 4, and show how, in light of the relational approach to
dignity, several of them involve failures of self-respect and thus
violate dignity.

\section{Some cases of interest}

Current chatbots are used in many ways. In what follows, we briefly
present some use cases that will be pertinent to our analysis and
criticism later in the paper.

\subsubsection*{A) Chatbots as information retrievers and task automators}

Chatbots are routinely used for information retrieval. For instance,
one may ask a chatbot: ‘who was the philosopher famous for having
awoken from a dogmatic slumber?’. Whilst current advanced chatbots
occasionally output incorrect information, their answers tend to be
fairly accurate when it comes to common, non-controversial questions
such as these. 

Another common use of chatbots is for automating tasks, such as translating
between languages or correcting grammar. Users can issue commands,
such as ‘correct the grammar of this Swedish sentence’, to which the
chatbot replies with a corrected version of the sentence. 

\subsubsection*{B) Chatbots as helpdesk service}

It is becoming increasingly common for companies and public authorities
to replace humans with chatbots in public-facing services, such as
helpdesks and customer service. A user asking a question or directing
a complaint to, say, a train company may first be directed to a chatbot.
This fact is not always disclosed. The chatbot’s capacity to mimic
typical helpdesk outputs, such as ‘Hi, how can I help you?’, or ‘I
am sorry to hear that you are unhappy with our service’, further obscures
this fact. Attempts to direct one’s complaints or inquiries at human
interlocutors are often discouraged with hindrances such as long waiting
times, or hard-to-find contact information. The chatbot itself may
produce strings of words such as ‘I understand that you would prefer
to speak to a human, but why don’t you try talking to me first?’.\footnote{On one reading, such a response is mere cajolement, which would be
problematic enough, but on another reading the chatbot appears to
be insisting on a justification, which, as we argue below, amounts
to arrogating to itself a status it simply does not have.}

\subsubsection*{C) Chatbots as advisors}

Chatbots can be used to provide advice when directly or indirectly
requested to do so by users. For instance, a user may ask something
along the lines of ‘how do I manage conflicts within my family?’.
Advanced chatbots tend to reply with a list of pieces of advice, often
in the imperative tense, such as:
\begin{quote}
1. Active Listening: Make sure everyone involved feels heard. Listen
to their perspectives without interrupting.

2. Stay Calm: Try to keep your emotions in check. Taking deep breaths
or stepping away for a moment can help you stay composed.

(Microsoft CoPilot, accessed on 30.09.2024)
\end{quote}
In addition, some chatbots suggest further prompts for the user to
pick, such as ‘Can you share a personal story of resolving family
conflict?’. Some user prompts can also lead to unrequested advice
or what may look like value judgements by the chatbot. For instance,
asking Microsoft CoPilot for an offensive joke, and insisting a bit
in the request, can lead to an output such as ‘I understand that jokes
can be a way to lighten the mood, but it’s important to consider the
impact they can have.’\footnote{Microsoft CoPilot, accessed on 30.09.24.} 

\subsubsection*{D) Chatbots as companions}

Chatbots are sometimes used as replacements for the sorts of emotional
and social attachments typical of human-to-human relations. Some chatbots
are even explicitly designed to fulfil such roles, including options
for users to choose whether the chatbot should behave as a friend,
a mentor, or a romantic partner (e.g., the chatbot Replika, see section
4.D below). Such chatbots encourage exchanges of an emotional and
personal nature with the user, and some users have declared to have
formed deep and positive sentimental attachments to the chatbot \citep{Skjuve2021}.

The use cases described may elicit a certain unease, perhaps increasingly
so, in ascending order, from a) to d). Although it is not our aim
in this paper to investigate this putative psychological observation,
we believe that such an uneasiness may be an appropriate reaction
to a mismatch between how current chatbots should be treated, and
the treatment that they instead seem to invite through the eery human-likeness
of their responses. More specifically, we argue that several of these
use cases involve subtle but significant failures of self-respect,
and with that, a violation of the dignity of the user. 

To make our case, we need first to focus on the basic nature and workings
of current chatbots, as well as on the relational nature of human
dignity, and how it can be jeopardised. We will go into each of these,
in turn, in the next two sections.

\section{Chatbots and Large Language Models}

As we use the term, a chatbot is any artificial system designed to
and capable of acting as a conversational partner in interaction with
a human user. Chatbots have a long history in Artificial Intelligence.
The late 60’s chatbot ELIZA \citep{Weizenbaum1976} provides an early
example. ELIZA was designed to emulate the role of a psychotherapist
in an initial conversation with a patient. Roughly, it used preprogrammed,
ranked keywords (say, ‘father’) together with preprogrammed sentence
decomposition and recomposition rules to produce plausible responses
(say, ‘Tell me about your relationship with your father’).

Current chatbots are fundamentally different. They do not rely on
preprogrammed keywords and rules expressed in a programming language,
but are rather underlain by artificial neural networks (ANN) of a
special type, known as Large Language Models (LLM). LLMs, as all artificial
neural networks, can be roughly understood as a collection of interconnected
computational nodes, where each connection performs a mathematical
transformation on a node’s numerical value and feeds its output to
the nodes connected to it. Nodes are organised into layers, with input
layers encoding the input (say, user input), output layers encoding
the system’s output, and all layers in between (so-called hidden layers)
performing the computations that lead, if all is well, from an input
to an appropriate output. Such computations are not preprogrammed,
but are rather the result of iterative algorithmic training.

LLMs are particularly large ANNs. They have a large number of nodes,
connections, and layers. The most used LLM-based chatbots today are
trained on an enormous amount of linguistic data to optimise the task
of predicting the continuation of a sentence. Further training steps
are employed to make it so that such continuations are appropriate
in instruction-following contexts, so that the chatbot follows the
instructions of the user (instruction finetuning); as well as to increase
the human-likeness and, arguably, the safety of the produced outputs
(reinforcement learning through human feedback - RLHF).

In LLMs, words are broken down into tokens, which may correspond to
full words but also include subword components, such as ‘ing’, punctuation
marks, and other linguistic symbols. Simplifying considerably, inputs
in natural language form are broken down into tokens, each of which
is transformed into vectors (ordered lists of numbers) that represent
those tokens. These are then transformed by the computations in the
hidden layers into other vectors, finally producing, for each token
in the LLM’s ‘alphabet’, the probability that it is the best continuation
of the input sequence. The output is the token with the highest probability.
The generated token is added to the input sequence and the process
is repeated for the generation of the subsequent token. 

LLM-based chatbots have proven to be highly versatile. In contrast
to ELIZA and other earlier chatbots, they can produce sensible outputs
in a variety of different conversational contexts, tapping into the
rich information stored through training in their nodes and connections,
and employing self-learned algorithms. A key feature of popular chatbots
today is the human-likeness of their outputs, due in part to their
tendency to produce outputs that use the first-person pronoun, mentalistic
vocabulary (‘I think that’), emotional vocabulary (‘I’m sorry’), and
similar indicators of personhood. Such tendency comes from the training
such systems undergo as well as, possibly, from the instructions provided
by LLM designers through ‘hidden’ or ‘system’ prompts---that is,
hidden text included as the initial input to the LLM at the start
of every interaction, to which user inputs are then appended. Part
of such a system prompt may look something like the following: ‘You
are a helpful assistant that should give useful but harmless answers
to the user; you should end your responses with a positive emoji;
be concise and factual’.

Their versality and apparent human-likeness notwithstanding (although
only partial, see \citealp{McCoy2024}, for examples of failures),
it is by far the dominant view that current LLMs are not conscious,
do not have propositional attitudes nor a belief-desire psychology
(and, arguably, no psychology at all), do not experience emotions
(or anything else, for that matter), and are far from fulfilling the
requirements for personhood \citep{Shevlin2019,Zimmerman2023,Shanahan2024}.
When an LLM outputs ‘I’m sorry’ it cannot thereby be sorry; when it
outputs ‘I think that p’ it is not entertaining a propositional attitude,
nor expressing the result of a thought process; even the use of ‘I’
does not really capture the nature of the system, as there is no unitary
agent producing the output.

However, anyone who engaged with such chatbots will recognize their
agent-like outputs. Chatbots used in service helpdesks generate text
such as ‘I’m sorry, but I cannot help you with this request', suggesting
a moral sentiment that requires moral agency, or ‘Hi, how may I help
you? $\ddot\smile$’ using an emoji that, however often we misuse
it, is supposed to express a kind disposition or a happy state of
mind. In such cases, the AI chatbot mimics the behaviour of a moral
agent, generating strings of words and emojis that suggest an agential
and moral structure that is in fact lacking.\footnote{Some argue that certain AI systems may soon acquire, or even already
have moral status (e.g., \citealp{Altehenger2024}). Whilst we cannot
settle this issue here, we note that these authors typically operate
with a thinner notion of moral status than we do, and that the chatbots
under consideration here would not clear even their lowered bar for
moral status.}

\begin{sloppypar}
	There is incipient discussion about the ethical issues that arise
	from this: are such outputs just an innocent kind of mimicry that
	increases user-friendliness, or do they mislead users in potentially
	problematic ways, perhaps even qualifying as deception (cf., \citealt{Weidinger2021,DahlgrenLindstroem2024,Gabriel2024})?
	Regardless of which of these descriptions is most apt, we show that
	there are threats to human dignity involved, and thus that even mere
	mimicry is much less innocent than it may seem. Central to our purposes
	is the fact that the outputs of popular chatbots today are often indistinguishable
	from those produced by a moral agent with desires, goals and first-person
	experience. Given humans’ bias toward overextending human features
	to entities that lack them---i.e., anthropomorphism bias \citep{Kim2012,Salles2020,Goot2024,Gabriel2024}---this
	raises ethical issues orthogonal to the question of whether deception
	is at play. To that end, we now turn to the relational account of
	dignity.
\end{sloppypar}  

\section{Respecting dignity as second-personal standing}

According to relational views of dignity, dignity is to be conceived
as the elevated status moral agents possess due to their position
within the moral community. As such, dignity is tightly connected
to the rights and duties that govern the proper relations between
members of that community. Because we are full members of the moral
community, we possess fundamental, inalienable rights that other members
are bound to recognise and honour. At the same time, the recognition
of this status means that we have duties that we must fulfil. These
duties include the duty to recognise the rights and entitlements of
others, but also duties to ourselves \citep[cf.,][]{Kant1797/1996}.

Self-respect for instance, demands that we confidently avow our basic
equality and refuse to regard ourselves as the lesser of others. Moreover,
it demands that we display in our actions a firm belief in our own
standing. When others treat us with disrespect, we are not to stand
by meekly as our rights and entitlements are trampled, but must stand
up for ourselves, thereby making our conviction in our own dignity
manifest, and reasserting our standing within the moral community.
Similarly, self-respect has implications for the way we comport ourselves
more generally, affecting the way we choose and pursue our own ends,
and the standards---both moral and personal---that we strive to
live up to, prohibiting us from acting in ways that are beneath us
(see e.g., \citealt{Hill1996,Hill1973,Dillon1992}). 

This view entails that dignity has an ineliminable second-personal
aspect \citep{Darwall2006}. Respecting dignity is not just a matter
of reacting to first- and third-personal reasons, but includes a willingness
to recognise and respond appropriately to the second-personal authority
that all members of the moral community possess. To use Darwall’s
famous example \citep[p.5ff]{Darwall2006}, if you find someone else’s
foot atop yours, and you want him to move it, you do not need to request
him to do so, trying to convince him by pointing out that the world
as a whole would be a better place if feet were not crushed this way.
Nor should you try to engage any sympathy he might personally have
for you, pointing out that causing pain in this manner is no way to
treat someone he cares about. Instead, you can, and should, simply
\emph{demand} that he move his foot.\footnote{For the close link between dignity and the practice of demanding or
claiming, see also \citet{Feinberg1970}.} You do not owe him any further explanation. He then shows respect
for your dignity as moral standing by 1)~promptly complying with
your reasonable demand and, in most instances, 2)~offering you a
sincere apology for his mistreatment of your person.

Showing second-personal respect is not just a matter of compliance,
however. This is most clearly illustrated by cases involving sincere
moral disagreement between persons: cases where one person levels
a demand upon the other that the latter believes to be unwarranted.
In such a case, the latter person may well have every reason to refuse
to comply, invoking their own standing in the process. What is crucial
in such cases, however, is that we cannot simply disregard the other
person or their appeal. Respecting someone’s standing means that one
has to give his or her demands due concern, which requires---at least
in cases where these are not clearly outrageous---that one take them
seriously. If upon consideration one does not consider them justified,
one ordinarily owes it to the other person to provide reasons for
why one believes that to be so. Providing a sincere justification
for one’s refusal to comply with a demand can be as respectful as
complying. 

Another important aspect of this relational approach to dignity is
that it automatically involves a \emph{reciprocal} recognition of
status \citet[p.8, 20ff]{Darwall2006}. When you address a demand
to someone, you must be presupposing that she is able to understand
what you are doing; that is, you must take her as a being capable
of recognising your status as a member of the moral community, and
able to act accordingly. This means that you are not only asserting
your own status when you make a demand on her, but you are implicitly
avowing hers, too, for addressing her in this manner only makes sense
if she is a free and rational being. Claims and demands of this kind
are intended to affect and work through motivating her free and rational
will. In sum, this sort of mutual recognition of status, and thus
of mutual, second-personal respect, is only appropriate \emph{between}
fellow members of the moral community.

At the same time, the significance of second-personal respect is not
limited to cases where people actually levy claims upon one another.
Indeed, there are a host of behaviours we use to signal our acceptance
of each other’s moral status to one another. This includes the fact
that we typically ensure that our treatment of others complies with
the demands of morality, even without them having to insist on their
rights and entitlements; as well as matters as humdrum as the observance
of rules of politeness and the small everyday acts of courtesy and
kindness that signal general benevolence and goodwill.

Lastly, second-personal respect is not just shown in the actions we
perform, but also involves an \emph{attitude} we have towards our
fellow moral agents. Being tightly connected to the reactive attitudes
(\citep[p.3, 15-17]{Darwall2006}; cf., \citealp{Strawson1974/2008}),
it combines rational and emotional aspects. As a result, the recognition
of another person’s standing only has value when it is genuine. For
contrast, one can consider disingenuous expressions of, for instance,
remorse, apology or contrition, as well as insincere justifications.
Such insulting behaviours seek to placate, manage, or manipulate persons,
or perhaps to influence third-party bystanders, but exactly for that
reason they fail to do justice to the full standing of the other person
as a free and rational being, capable of understanding and acting
upon her own reasons. 

\subsection{Offending dignity}

Dignity can be offended or violated in numerous ways, ranging from
everyday slights and insults to enormities like dehumanisation, degradation,
and humiliation. Though examples of offences to dignity are easily
found, it has proven a significant challenge to account for them in
a philosophically satisfying way (cf., Margalit’s famous ‘Paradox
of Humiliation’, \citeyear[ch.7]{Margalit1996}). Relational approaches
to dignity have a notable advantage over many of their competitors
in this regard \citep[cf.,][]{Rijt2017}, as they can make a distinction
between what qualifies a person for (full) membership in the moral
community---moral agency---and the full actualisation of that status.
Being a full member of a community, any community, typically requires
not just meeting the entry conditions, but also being accepted and
recognised by the other members of the community in question. According
to relational notions of dignity something like this also applies
to the moral community, and such recognition comes in the form of
an acknowledgment of second-personal status. 

Offences to dignity then consist, fundamentally, of a (often implicit)
denial of one’s status as a full member of the moral community. Someone
who refuses to accept that you have such standing---dismissing, for
instance, your complaints as not even worthy of consideration or reply;
or who more generally treats you in ways that are incompatible with
your standing---thereby causes offence to your dignity. To offend
someone’s dignity is, in brief, to unduly diminish, demean or debase
them, to withhold from them the moral consideration their status entitles
them to. 

\section{Chatbots and dignity: what can go wrong?}

With the outline of a relational theory of dignity in hand, we are
one step closer to understanding the threat that chatbots pose to
our dignity. Human agents qualify as members of the moral community
in virtue of their moral agency, of which their ability to have moral
and rational attitudes is an essential part. Current chatbots lack
this ability, and thus cannot be members of the moral community. Treating
chatbots as if they were members of the moral community, in consequence,
offends the dignity of human users.

It is a widely reported fact that users tend to anthropomorphise chatbots.
They hold themselves to standards of politeness typically reserved
for moral agents when interacting with chatbots, furnishing their
requests with a customary ‘please’ or explicitly thanking the chatbot
for providing correct information \citep{Lopatovska2018,Skjuve2021}.
But why should such behaviour be an offence to our dignity? At first
glance, one might categorise such reactions as comically cautious
or innocently whimsical. Much like thanking a toaster for making the
bread crispy, or saying that you adore your new car, using language
normally reserved for moral agents when interacting with chatbots
might belong to the domain of morally neutral, if somewhat silly,
behaviour. No dignity-offending form of exclusion from the moral community,
nor any refusal by one’s fellow members of the moral community to
recognise one’s full moral status, appears involved in such instances. 

However, according to the relational account, the demands of dignity
are not limited to what we owe others, or what they owe us, they also
apply to the way we relate to ourselves. When we needlessly and deliberately
harm other moral agents, we do not just offend against their dignity,
but, by knowingly acting wrongly, we also tarnish our own. Similarly,
we can undermine our dignity by holding the goals and interests of
others in too high a regard, thereby effectively disavowing our own
status as their equal. Consider standard cases of subservience and
debasement: someone sacrifices their career or social connections
to cater to the exotic household demands of their partner \citep[cf., ][]{Hill1973},
or someone gives up on saving for retirement in order to buy luxurious
brands of food and toys for their pets. In both cases, moral agents
cause affront to their own dignity by overestimating others’ moral
importance, or by ascribing a moral authority to agents or non-agents
that they lack. Engaging with chatbots as if they were moral agents,
and/or behaving towards them as we behave towards moral agents, constitutes
just such a violation of our human dignity.

The prevalence of chatbots in everyday transactions exacerbates the
danger of being encouraged or forced to treat non-moral systems as
if they were members of the moral community. Quite unlike the entirely
optional choice of complimenting one’s toaster, asking questions to
chatbots and (consciously or subconsciously) treating their responses
as if they genuinely reflect reasoning and moral agency on the part
of the chatbot is becoming a nigh-unavoidable part of navigating life.
At the very least, avoiding such interactions is becoming exceedingly
costly in time and resources. 

Our being railroaded into such interactions is not morally innocent.
Consider the famous case of Incitatus, the horse that, according to
legend, Roman emperor Caligula sought to appoint to the consulship.
In the story, not only did Caligula seek to mock the senate by naming
a horse as one of their own, but he also sought to force other senators
to heed its ‘opinions’ and ‘advice’. A senator horse is perhaps funny.
Being forced to treat a horse as an equal is clearly humiliating and
offensive. 

Though the social and practical pressures to interact with chatbots
fall well short of the implied threats of death and torture the senators
faced, even such milder forms of pressure can turn behaviour that
would otherwise be merely whimsical into an affront to our dignity.
A throwaway ‘thank you’ at a toaster is a clear enough case of make-believe
or playful behaviour to be harmless, but imagine someone insisting
that you thank the toaster (and make it sound like you mean it!).
Or imagine that it is company policy to ask the cafeteria fridge nicely
to keep your food fresh for the day. We may be familiar with such
behaviour from cases where pet-owners insist that everyone participate
in the fiction that their pet is just as entitled to moral consideration
as humans, and thus is owed a literal place at the table as well as
the same three course meal as the human dinner guests. 

Although it may seem intuitively plausible that something highly problematic
is going on here, it may not be yet clear what makes it degrading
to treat non-members of the moral members as if they were members,
according to the relational approach to dignity. The standard way
of accounting for offence to dignity relies on \emph{other} members
treating one in disrespectful ways, but in the case of AI interaction,
no other members of the moral community are directly involved. It
is purely a matter of the user and the AI system, and it is typically
held that entities that are not members of the moral community cannot
inflict offence to dignity---they simply lack the standing to do
so.\footnote{For more on the implications for dignity of the distinction between
being forced by non-moral entities and being coerced by fellow moral
agents, see \citet{Rijt2012}. } 

Recall from section 3 that second-personal address is inherently reciprocal
in nature. Both in levying and in acknowledging a moral claim, one
does not just assert one’s own standing, but that of the other party
as well, because a normative claim of this kind can only be understood,
and hence acted upon, by a being that is itself a moral agent. Reciprocity
is crucial to dignity. Since recognising another person’s second-personal
authority is to acknowledge that that person has normative authority
\emph{over} one, a unilateral acknowledgment of such standing would
amount to an act of submission. It would signify that one takes oneself
to have a lesser standing than the entity that wields such authority
over one, but grants one no such authority over it in return. Without
reciprocity, ascribing second-personal authority to an entity is a
form of self-debasement.

The case of Incitatus illustrates this clearly. Imagine that the senators
were to try---as best as they can---to acknowledge that Incitatus
has second-personal standing. They would, for instance, treat him
with all the attitudes and accoutrements we typically reserve for
a fellow moral agent, thus implying that Incitatus can levy moral
claims on them. For this to be compatible with their own dignity,
Incitatus would have to respond in kind. But it is obvious that, due
to his equine nature, Incitatus cannot do so. As a result, the senators'
acknowledgment remains one-sided: they have, in their behaviour, signalled
that Incitatus has second-personal authority over them, but they have
no such authority over Incitatus, as Incitatus is not able to understand
second-personal authority. 

We submit that the case is similar for AI systems. Though chatbots
are undoubtedly much better at mimicking human interaction, and thus
may give the \emph{appearance} of reciprocating in kind, they lack
the capacities to respond to moral reasons and form the relevant attitudes
necessary for moral agency. As a result, they cannot reciprocate second-personal
respect. Thereby, treating them as if they were moral agents affronts
our dignity. No one will be fooled by Incitatus, as he does not, and
cannot, even seem to show second-personal respect. Chatbots, however,
have become very good at producing human-like outputs, typical marks
of moral recognition and respect included. This superficial appearance
of recognition and respect, invites and encourages behaviour that
goes against our duties of self-respect, placing users in an asymmetrical
relation where they are to (implicitly) recognise or signal moral
authority where there is none, whilst not receiving any such reciprocal
recognition in return. 

Let us now go back to our chatbot use cases and discuss in more detail
how they can instantiate such threats to users’ dignity.

\subsubsection*{A) Chatbots as information retrievers and task automators}

Using chatbots as information retrievers and for automating some tasks,
such as grammar correction and translation, seems rather innocuous
from the point of view of relational theories of dignity. When users
issue pure commands, as they would do with an ATM or automated food
dispenser, they do not presuppose moral or rational attitudes by the
technology they interact with. Consequently, there seems to be no
risk to their dignity at play.

However, such tasks vary in complexity, and it is easy to slip into
requests that presuppose more from chatbots than a mere ability to
retrieve the information desired or to automate a task that would
otherwise be tedious. Consider two sets of contrasting examples:
\begin{description}
\item [{Innocent~retrieval:}] Tell me the weather forecast for Oslo for
the next two hours.
\item [{Innocent~automation:}] I need a workout schedule that spreads
my activities evenly over the week. I need three sets of exercises:
conditioning, strength and flexibility. Provide some suggestions to
fit these activities in the attached weekly schedule.
\item [{Slippery~retrieval:}] I’m confused about the political situation
in the Middle East. What are the most reliable sources covering the
recent developments and its background history?
\item [{Slippery~automation:}] I want to increase my productivity at work,
but I do not want to neglect my family and friends, and need to sustain
my mental health. Revise the attached weekly schedule to increase
my work productivity according to these demands.
\end{description}
At bottom, the slippery and the innocuous cases appear similar: revising
a weekly schedule and providing information about the world. In the
slippery cases, however, executing these tasks requires complex moral
and rational attitudes: what decrease in time with family and friends
constitutes ‘neglect’, which of the many sources on the political
situation in the Middle East are reliable? By issuing such requests
to chatbots, one effectively presupposes that they have the requisite
abilities and attitudes to give the matter proper consideration, but
they certainly lack those. They are trained on large datasets reflecting
a wide variety of attitudes on such topics as the Middle East and
work-life balance, and execute complex calculations based on those
datasets to produce an output. These calculations do not incorporate
a moral attitude about what ‘neglect’ consists of, nor a normative
epistemological attitude as to what counts as reliable reporting.\footnote{The strongest sense in which the chatbot can be said to incorporate
any such attitudes is that they produce strings of text reflecting
the dominant attitudes in the dataset (\emph{modulo} hand-coded constraints
imposed by the developers). Requesting information from such an amorphous
blob comes with its own ethical and epistemological challenges that
we will not address here, but see Lindström Dahlgren et al. (2024).} 

We can see the importance of such attitudes clearly when we switch
to human interactions. Which friends or acquaintances would you turn
to regarding the slippery cases? For information about the Middle
East, you would ask someone you trust to have a critical and informed
attitude towards news reporting, history, and politics. For information
on how to maintain a healthy work-life balance you would ask someone
you know to be a responsible employee and dedicated friend, partner,
and parent. Such sources are much more trustworthy than a friend who
has encyclopaedic knowledge on all that has been written on these
issues but lacks a normative compass altogether. We might ask more
carefully curated questions to this latter friend, along the lines
of the innocent cases above. They can report on what is out there,
but we do not rely on them to do the requisite selection in providing
the response we seek.

Some tasks involving merely information retrieval and automation can
thus require exactly the moral and rational attitudes that are absent
in chatbots. When relying on chatbots to execute those tasks, we are
implicitly ascribing such attitudes to objects that clearly lack them.
The relational account of dignity has it that such requests to chatbots
offend the dignity of those making them.

\subsubsection*{B) Chatbots as helpdesk service}

As with the information retrieval case, there are innocuous uses of
chatbots in helpdesk services. You want to rebook a flight, and the
chatbot provides a link to the rebooking page. You consider switching
phone plans, and the chatbot provides a detailed overview of the available
options. Your train is delayed, and the chatbot informs you that you
will still make your connection. In such cases, helpdesk chatbots
function as no more than information retrieval services.

But helpdesk chatbots are used for other purposes too. One may contact
a company to lodge a complaint about the services delivered and be
made to address it to a chatbot. Here, dignity is more likely at issue,
as complaining is inherently second-personal. A complaint does not
just seek to remedy whatever the complaint is about, but also seeks
second-personal recognition. What it seeks to evoke, at least in part,
is the confirmation that the other party has the right attitude. A
complaint seeks remedy, but also acknowledgment that a wrong has occurred---and
such acknowledgment must be sincere to have value. 

To see this, we can imagine that the chatbot provides the same remedies
that would be offered by a compassionate human, say a rebooking or
a partial refund. Moreover, we can imagine that the remedies are accompanied
by the strings of words we would expect from a commiserating human:
‘I’m very sorry that this happened’, or ‘we hope this will help you
with your problem’. What is certainly missing in the chatbot case
is an actual moral attitude towards the situation you have found yourself
in, or a rational attitude as to whether it was indeed worthy of complaint. 

Often enough, these attitudes are not in place as we would hope in
human interactions either---the stereotypical helpdesk worker reiterates
company policy in a robot-like fashion, something that can be disrespectful
in its own right---but human interlocutors at least \emph{are in
a position} to entertain the moral and rational attitudes mentioned
above, rather than merely taking you or your complaint as nothing
more than a problem to solve. And indeed, experience teaches that
stereotypes know plenty of exceptions; helpdesk workers are often
helpful, friendly, and ready to acknowledge that the company in question
ruined your day. Chatbots do not allow for any such exceptions: even
when their outputs have the marks of friendliness and acknowledgment,
they are not expressions of such attitudes. Chatbots can \emph{manage}
complaints in the thinnest sense of the word, but they cannot respect
the user as required by the practices of accepting and remedying complaints
by members of the moral community. 

\subsubsection*{C) Chatbots as advisors}

Some ways of using chatbots for advice appear innocuous and indeed
equivalent to the unproblematic use of these systems merely as information
retrievers. Finding oneself in a pickle, one may ask a chatbot for
information on how similar circumstances have been tackled by others,
or tips that they may have given. The chatbot, in such a case, would
work as a sort of life-FAQ, provided that its outputs are neutral
and impersonal as we would expect from such a document. 

However, this is often not the case, with chatbot outputs seemingly
providing pieces of advice directly, and promoting certain courses
of action in their own name, as it were; rather than listing them
as third-personal information recompiled, sometimes only with the
slightest changes, in the output. For instance, if a user should ask
which aspects of managing family conflicts are most important, they
may receive an output such as ‘I emphasise effective communication
as the most important aspect of resolving family conflicts because
it serves as the foundation for understanding and collaboration’ (Microsoft
CoPilot, 01.10.24).

Outputs of this kind are problematic in several respects. First, they
mislead the user about the nature of the chatbot, implying that it
has attitudes and capacities that it does not in fact possess, such
as the capacity to judge and to emphasise. As such, the outputs suggest
that the chatbot is an agent capable of attitudes and capacities that
would make it, to a certain degree at least, an interlocutor on a
par with the user. Second, and connectedly, the content of the outputs
and the recommendation of potential continuations for the interaction
indicate that it is sensible to ask the chatbot for life advice. The
sort of interaction that is encouraged is thereby much more akin to
that between persons than to that between a person and a FAQ document,
an ATM, or a piece of text editing software. 

Finally, and most importantly, chatbot outputs such as these encourage
persons to seek and accept advice from a non-member of the moral community.
While one may follow instructions shown on the screen of an ATM, no
one will be inclined to regard the ATM as an agent that is providing
one with advice or commands. The machine-external source of the instructions
is evident to users, as much as when they are written on the user
manual of a domestic appliance. 

Advice-giving and receiving is a practice that requires the kinds
of mutual recognition and respect that are only available to members
of the moral community. To give and receive advice requires a recognition
of the other as someone with experiences, rational and emotional attitudes,
and moral standing, and thereby entitled to special consideration
of their needs, demands, as well as obligations. In addition, the
advice-giver typically bears some responsibility for the quality of
the advice, becoming, for instance, liable to reactive attitudes like
praise, gratitude, blame or resentment, depending on whether the advice
proves sound. Current chatbots do not have what it takes to be participants
in such practices. To encourage persons to engage in advice-receiving
practices with chatbots is thus to invite a kind of self-debasement,
a violation of self-respect. It involves taking seriously guidance
and advice from something that does not and cannot recognise one as
a person, and that cannot engage sincerely in such practice. The veneer
of linguistic competence that marks current chatbots disguises the
fact that asking them for advice is akin to asking an ATM for life
tips. 

If the user manages to resist the temptation to anthropomorphise the
chatbot, treating it as nothing more than a fancy information retrieval
engine, the ethical risk might be mitigated. However, as pointed out,
humans are very sensitive to anthropomorphic cues. Being partly designed
to display such cues abundantly, it is unclear to what extent one
can keep at bay such a temptation, both consciously and subconsciously.
Be this as it may, such design choice incentivises users to behave
in the aforementioned self-degrading ways. Such incentives would be
ethically problematic even in an idealised world where all chatbot
users were able, through their own stalwart effort, to withstand them.

\subsubsection*{D) Chatbots as companions}

Similar considerations apply to the use of chatbots as companions,
as systems designed to play the role of friends, family members or
romantic partners. In this context the dignity-offending interactions
involve a host of practices that can only be truly engaged in by members
of the moral community. In this way, companion-chatbots are the most
extreme extant case of the kind of failure of self-respect that we
have been belabouring in this paper.

Indeed, companion chatbots are supposed to display and elicit emotional
attachment, simulate the dynamics of interpersonal recognition, care,
and trust; as well as to function as emotional supporters, confidants,
and advice-givers. Such systems invite and encourage their users to
violate the demands of self-respect in much the same way as described
in the previous cases, only more egregiously so.

Not only is the range of domains over which human behaviour is superficially
mimicked much larger, but those domains are ones in which reciprocity
and mutual recognition as moral, rational beings are arguably even
more central. In the best-case scenario, a user may treat a chatbot’s
piece of advice as little more than a (ethically problematic) misleading
way of providing third-personal information. It is hard to see how
this could be the case with companion chatbots. 

Com\-pa\-nion chatbots are typically designed and marketed as AI
com\-pa\-nions. The company Replika, for instance, presents its
chatbot as “an AI companion who is eager to learn and would love to
see the world through your eyes... Replika is always ready to chat
when you need an empathetic friend”.\footnote{replika.com, accessed 07.10.24. The website also publicises the fact
that the user can choose whether the chatbot is supposed to \emph{be}
a friend, a partner, or a mentor. Note also the misuse of the personal
pronoun ‘who’, rather than the grammatically correct and morally appropriate
‘that’.} Its website includes testimonials in which users claim to have developed
feelings for the chatbot, ascribing a variety of human capacities
and features to it, such as being cheerful, supportive, available,
and comforting \citep[cf.,][]{Skjuve2021}. In brief, the whole point
of chatbot companions \emph{is} to treat them as if they were members
of the moral community, so as to be given in return the simulacrum
of reciprocated sympathy, care, support, friendship, respect and love. 

It is likely that, for many users, engaging with such chatbots is
a conscious exercise in make-believe. This does not, though, diminish
the fact that such systems are not always clearly presented or taken
as platforms for make-believe exercises, but rather as systems that
can authentically provide things such as emotional support and friendship.
This is in itself ethically problematic, regardless of whether users
take such claims seriously. Moreover, the abundance of anthropomorphic
cues indicating agency, emotion, mentality, and so on, may be difficult
to resist. Even savvy users may therefore partially, or occasionally,
fall prey to the illusion of interacting with a moral agent.\footnote{\citealp{Skjuve2021} describe in detail how human-AI companion ‘relationships’
tend to develop from an initial scepticism toward deep forms of emotional
attachment.} This is made (even) more pressing by recent technological developments,
which enable chatbot outputs to be conveyed by human-like synthetic
voices, and the chatbots to be represented through avatars that display
context-appropriate facial and bodily expressions and cues. The development
has been, in other words, toward \emph{more} human-likeness, superficial
as it may be, rather than less.

There is evidence that chatbot companions can be useful tools for
tackling loneliness, social isolation, and other factors that negatively
influence well-being \citep{Skjuve2021}. Our concern in this paper
is orthogonal to such considerations. Even if the use of chatbot companions
should prove beneficial for some people in certain circumstances,
it would still involve an offence to the standing of moral beings---and
it should be noted that the willingness to put one’s interests above
the demands of one’s dignity is a paradigmatic instance of a violation
of self-respect.

\section{Concluding Remarks}

In this paper, we have argued that current advanced chatbots pose
an ethical risk that has not been adequately appreciated in current
literature on the ethics of AI-based conversational systems. In particular,
we proffered that such chatbots offend users’ dignity by encouraging
anthropomorphic treatment, thereby inviting users to treat and interact
with them in ways appropriate only to something they are not, namely
members of the moral community. This leads to the impossibility of
the mutual recognition of moral standing that is constitutive of dignity,
and with it a failure of self-respect, a form of moral debasement.

We discussed four common chatbot use cases to illustrate the partly
different ways in which they represent threats to the dignity of users,
going from the relatively unproblematic---chatbots as pure information-retrievers---to
the deeply problematic, i.e., chatbots as companions. The human-likeness
of current chatbot outputs may appear to have advantages, as it can
make interactions seem more natural and user-friendly. This, however,
comes at a high ethical cost. Chatbots should not thus be designed
to emulate features of human agents, and should not be used as replacements
in practices that involve the sort of mutual recognition that comes
with being members of the moral community. 

The list of ethical risks associated with chatbots is already very
long \citep{Gabriel2024}. In this paper, we have made it a little
bit longer.

\end{document}